\documentclass[
reprint,
longbibliography,
superscriptaddress,
 amsmath,amssymb,
 aps,
 pre,
floatfix,
]{revtex4-2}

\usepackage{amsmath,amssymb}
\usepackage{physics}
\usepackage{graphicx, color}
\usepackage{dcolumn}
\usepackage{bm}
\usepackage{bbm}
\usepackage{hyperref}
\usepackage[dvipsnames]{xcolor}
\hypersetup{
setpagesize=false,
 bookmarksnumbered=true,%
 bookmarksopen=true,%
 colorlinks=true,%
 linkcolor=Blue,
 citecolor=Blue,
 urlcolor = Blue,
}

\newcommand{\RM}[1]{\mathrm{#1}}

\newcommand{\eq}[1]{Eq.~(\ref{#1})}
\newcommand{\fig}[1]{Fig.~\ref{#1}}

\newcommand{\p}{\partial}

\begin{document}

\preprint{APS/123-QED}

\title{Singular density correlations in chiral active fluids in three dimensions}

\author{Yuta Kuroda}
\email{kuroda@r.phys.nagoya-u.ac.jp}
\affiliation{%
 Department of Physics, Nagoya University, Nagoya 464-8602, Japan
}%
\affiliation{%
 Institute for Theoretical Physics IV, University of Stuttgart, Heisenbergstr.~3, 70569 Stuttgart, Germany
}%

\author{Takeshi Kawasaki}
\affiliation{%
 Department of Physics, Nagoya University, Nagoya 464-8602, Japan
}
\affiliation{
D3 Center, The University of Osaka, Toyonaka, Osaka 560-0043, Japan
}
\affiliation{
Department of Physics, The University of Osaka, Toyonaka, Osaka 560-0043, Japan
}

\author{Kunimasa Miyazaki}%
\affiliation{%
 Department of Physics, Nagoya University, Nagoya 464-8602, Japan
}%

\date{\today}

\begin{abstract}
We investigate density fluctuations in three-dimensional chiral active fluids by using a simple model of helical self-propelled particles.
Helical motion is generated by a constant angular velocity (or chiral torque) acting on the self-propelled force.
The chiral torque is assumed to have the same direction and magnitude for all particles.
Due to the helical nature of the particle motion, the system is generically anisotropic even when it is spatially homogeneous.
Numerical simulations demonstrate that the helicity induces an anisotropic pattern and a singularity in the static structure factor (the density correlation function in Fourier space) in the low-wavenumber limit.
Moreover, the system in the limit of infinite persistence time exhibits hyperuniformity in the direction perpendicular to the chiral torque, while giant density fluctuations emerge along the parallel direction.
We then construct a fluctuating hydrodynamic theory for the system to describe the singular behavior. 
A linear analysis of the resulting equations yields an analytical expression for the static structure factor, which qualitatively agrees with our numerical findings.
\end{abstract}

\maketitle

\section{Introduction}
Self-propelled particles have been widely used to model active matter and have revealed numerous novel phenomena arising from their nonequilibrium nature~\cite{Ramaswamy2010Anuual_Review, Bechinger2016RMP, Andreas2023annual}. 
Among various types of self-propelled particles, chiral self-propelled particles, whose motions violate mirror symmetry, have recently attracted significant attention~\cite{Lowen2016, Liebchen_2022}. 
In two dimensions, they typically exhibit circular trajectories.
Experimental realizations of such chiral self-propelling motions include asymmetric-shaped colloidal objects~\cite{Kummel2013PRL, Zhang:2020ux}, Janus particles with an asymmetric coating~\cite{Mano2017}, sperm cells~\cite{Ingmar2005Science}, malaria parasites~\cite{Patra:2022vd}, bacteria swimming near walls or surfaces~\cite{DiLuzio2005, Leonardo2011}, and light-driven walkers~\cite{Siebers2023}. 
In numerical simulations, such circular motion is often modeled by introducing chirality into active Brownian particles (ABP)~\cite{Lowen2016, Liebchen_2022}.
It has been reported that the chiral motions of constituents give rise to collective phenomena that are not observed in non-chiral counterparts. 
Examples include formation of vortices~\cite{Liebchen2017, Kruk2020PRE, Ventejou2021PRL, Liao2021SoftMatter}, absorbing phase transition~\cite{Li2019}, reentrant behavior of glassy slow dynamics~\cite{Debets2023PRL}, anomalous two-dimensional crystallization~\cite{kuroda2025PRR, jeong2025}, Hall-like currents~\cite{Siebers2024PNAS}, and self-reverting vorticity of clusters~\cite{Caprini:2024ab}.
Besides chiral self-propelled particles, active spinner systems, where particles rotate but do not self-propel, have also been studied intensively~\cite{Aragones:2016aa, Soni:2019aa, Tan:2022aa, Liebchen_2022}. 

In addition to the aforementioned examples, one of the most striking features of two-dimensional chiral self-propelled particles is the suppression of long-wavelength density fluctuations~\cite{Li2019, Huang2021, Zhang2022, kuroda2023_chiral}.
This phenomenon is known as hyperuniformity~\cite{TORQUATO, Lei_2025_review}.
Hyperuniformity is characterized by the vanishing of the static structure factor $S(q)$ in the small-wavenumber limit, with $q$ denoting the wavenumber.
In most cases, the asymptotic behavior is a power law as $S(q\rightarrow 0)\sim q^{\alpha}$ with $\alpha >0$. 
This feature is in contrast to non-chiral active fluids, which typically exhibit enhancement of density fluctuations (also known as giant number fluctuations:~GNF)~\cite{Ramaswamy2003EPL, Narayan2007Science, Zhang2010PNAS, Nishiguchi2017PRE, Iwasawa2021PRR, Chate2006PRL, Chate2008PRE, Ginelli2012PRL, Kawaguchi:2017aa}. 
Hyperuniformity has been observed in a variety of nonequilibrium dynamical systems~\cite{Lei_2025_review}, such as periodically driven colloidal suspensions near absorbing transition points~\cite{Weijs2015, Tjhung2015PRL, Hexner2015PRL}, center-of-mass-conserving systems~\cite{Hexner2017, Lei2019_pnas, Galliano2023PRL, ikeda2023, maire2025}, granular systems~\cite{maire2024, maire2025, maire2025granular}, non-reciprocal active robotic systems in chiral states~\cite{Chen2024PRL}, epithelial tissues with pulsating synchronization~\cite{li2025PNAS}, deterministic anti-aligning polar active fluids~\cite{boltz2024}, and the late-stages of spinodal decomposition described by the Cahn--Hilliard equation~\cite{Ma2017HU} and its active counterpart (active model B without noise)~\cite{zheng2023}.

While most studies have focused on two-dimensional systems, three-dimensional chiral active particles, whose motion becomes helical, are also of great interest.
Examples of such systems include sperm cells~\cite{Jikeli:2015aa}, choanoflagellates~\cite{Kirkegaard2016PRL}, and nematic liquid crystal droplets in surfactant solutions~\cite{kruger2016PRL, Yamamoto2017SoftMatter}, all of which are known to exhibit helical trajectories.
Theoretically, the single-particle dynamics of three-dimensional extensions of chiral active Brownian particles~\cite{Sevilla2016PRE, Pattanayak_2024}, as well as of helical swimmers driven by hydrodynamic interactions with the surrounding fluid~\cite{Burada2022PRE, Lisicki2018SoftMatter, Rode:2021aa}, have been investigated.
Yet, the collective behavior arising from helical motion has received relatively little attention, with the exception of a study on hydrodynamically interacting helical swimmers~\cite{Samatas2023PRL}.  

In this paper, we study density fluctuations in a collection of helical active swimmers using a simple particle model.
Helicity is introduced by adding a constant angular velocity (or torque) to the equation of motion for the self-propulsion in three-dimensional ABP, following the approach in Refs.~\cite{Sevilla2016PRE, Pattanayak_2024}.
In this model, the key parameters are the magnitude of the torque and the persistence time (or the rotational diffusion constant).
Each particle exhibits helical motion along the direction of the torque vector, resulting in anisotropic dynamics. 
The torque vector is assumed to be aligned in the same direction for all particles.
We perform extensive numerical simulations in the homogeneous fluid states and find that the density correlation function exhibits a singular behavior, characterized by a discontinuity in the low-wavenumber limit of the static structure factor.
In addition, we observe an anisotropic pattern in the static structure factor.
These properties are akin to those observed in other anisotropic nonequilibrium fluids, such as active particles with anisotropic self-propulsion~\cite{Adachi2022PRR, Nakano2024PRR, Adachi2024PRR} and in driven diffusive gases~\cite{Schmittmann1995, Schmittmann1998}.
In our system, density fluctuations are enhanced along the direction of the torque, while they are suppressed in the perpendicular directions.
In particular, in the limit of infinite persistence time, the system exhibits hyperuniformity in the directions perpendicular to the torque.
To understand these numerical observations, we construct a fluctuating hydrodynamic theory by a bottom-up approach and analytically calculate the static structure factor.
The resulting expression qualitatively captures the observed behaviors in the numerical simulation, although some quantitative discrepancies remain.

This paper is organized as follows.
In Sec.~\ref{sec2}, we present the details of the particle-based model for helical active swimmers.
Section~\ref{sec3} provides numerical results for the density fluctuations in the homogeneous fluid states.
To explain the qualitative features observed in the simulations, we develop a fluctuating hydrodynamic theory in Sec.~\ref{sec4}, and derive an analytical expression for the density correlation function via linear analysis.
Finally, Sec.~\ref{sec5} summarizes our results.

\begin{figure}[t]
\centering
  \includegraphics[width=9cm]{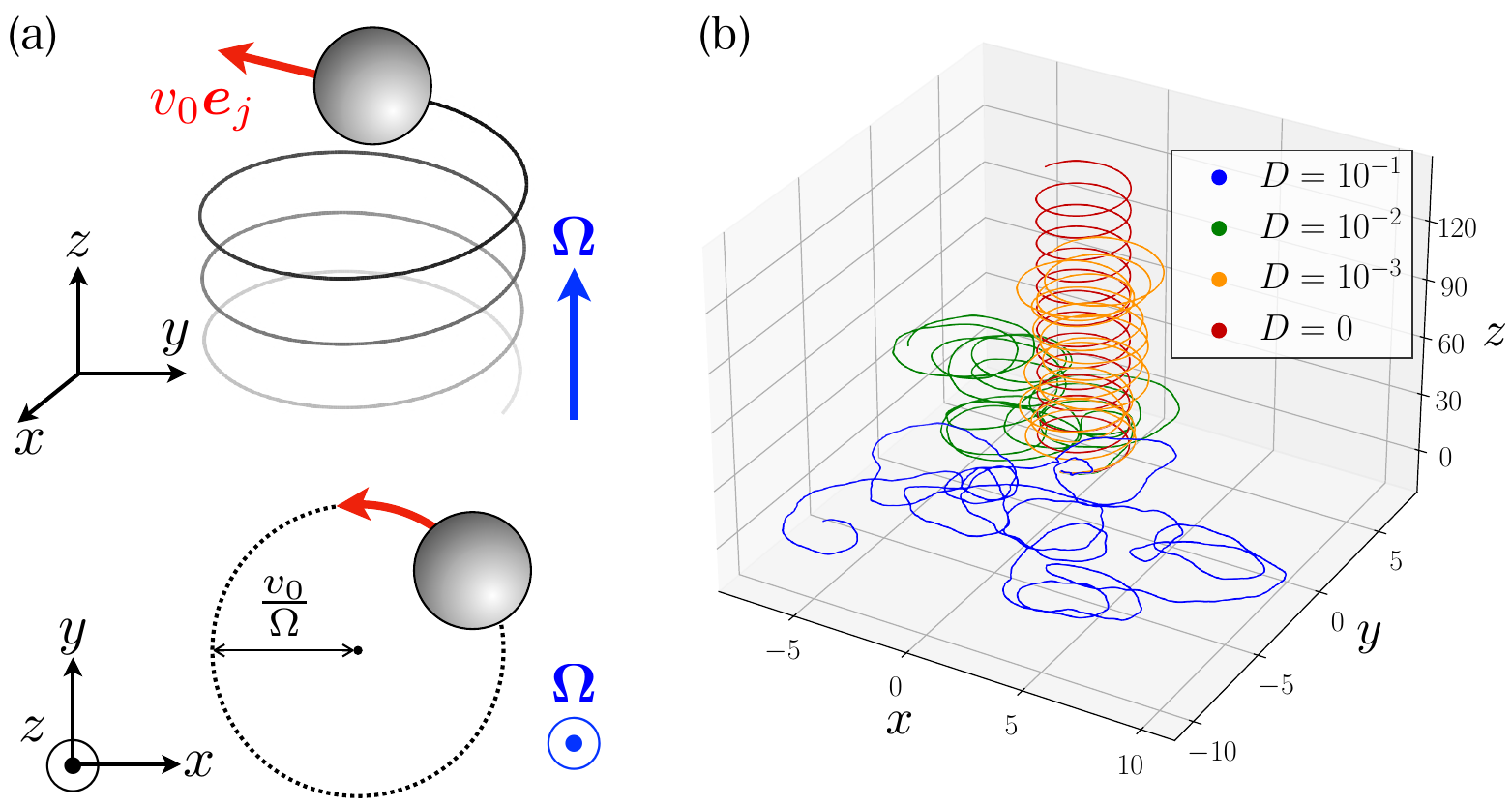}
  \caption{\label{fig1}(a)~Schematic illustration of the model of a helical self-propelled particle. The illustration shows the case of $\bm{\Omega} = (0, 0, \Omega)$ and zero noise. The top panel depicts the motion in three dimensions, and the bottom panel shows its projection onto the $x$-$y$ plane.
(b)~Typical trajectories of a single free particle for various rotational diffusion constants.
The chiral torque is fixed as $\bm{\Omega} = (0, 0, \Omega)$ with $\Omega = 0.4$.
Time is measured in units of $\tau = \sigma / v_0$.
}
\end{figure}
\section{Model}
\label{sec2}
We study a model of helical self-propelled particles~\cite{Sevilla2016PRE}, which we here refer to as 3D cABP, illustrated in \fig{fig1}(a).
The equation for the position of the $j$th particle is given by
\begin{align}
\dv{\bm r_j(t)}{t} = \mu\bm F_j(t) + v_0\bm e_j(t), \label{eq7-1} 
\end{align}
where $\mu$ and $v_0$ stand for the mobility and the constant self-propulsion speed, respectively.
The interaction force acting on the $j$th particle is given by $\bm F_j(t)  = - \sum_{k(\neq j)}\nabla _{j}U(\abs{\bm r _{j} - \bm r_{k} })$, where $U(r)$ is a short-range pairwise potential.
$\bm e_j=(e^{(x)}_j,e^{(y)}_j,e^{(z)}_j)$ is a three-dimensional unit vector and represents the direction of the self-propelled motion.
This unit vector can also be expressed by two polar angles, $\phi_j$ with respect to the $x$-axis (the azimuthal angle) and $\theta_j$ with respect to the $z$-axis (the zenith angle), as  $\bm e_j=(\sin\theta_j\cos\phi_j,\sin\theta_j\sin\phi_j,\cos\theta_j)$.
We assume the time evolution of the unit vector $\bm e_j(t)$ to obey
\begin{align}
\dv{ \bm e_j(t)}{t} = \qty(\bm \Omega + \sqrt{2D} \bm \eta_{j}(t) ) \overset{\RM{s}}{\times} \bm e_j(t). \label{eq7-2}
\end{align}
Here, $D$ denotes the rotational diffusion constant. 
The persistent time in three dimensions is given by $\tau_\RM{p}=1/(2D)$. 
The vector $\bm \Omega$ is a constant torque that generates helical motions. 
The unit vector $\bm e_{j}(t)$ is subjected to a Gaussian white noise $\bm \eta_{j}(t)$ with zero mean and the correlation $\expval*{\eta _{j}^{(\alpha)}(t)  \eta _{k}^{(\beta)}(t')} = \delta_{j,k}\delta_{\alpha,\beta}\delta(t-t')$, where the symbol $\expval{\cdots}$ denotes the ensemble average and the Greek indices denote Cartesian coordinates: $\alpha,\beta \in\{x,y,z\}$.
Note that the cross product in \eq{eq7-2}, $\overset{\RM{s}}{\times}$, is taken in the Stratonovich representation~\cite {gardiner}. 
This model is a simple extension of two-dimensional chiral active Brownian particles~\cite{Ma2017, Li2019, Ma2022, Debets2023PRL} to a three-dimensional case.
In the limit $\bm \Omega \rightarrow \bm 0$ at finite $D$, Eqs.~(\ref{eq7-1}) and (\ref{eq7-2}) reduce to active Brownian spheres or three-dimensional ABP~\cite{Wysocki_2014, Stenhammar2014soft_matter, Farage2015PRE, Siebert2017}. 
We remark that, in reality, each helical swimmer likely has a different direction and magnitude of the torque.
However, to focus on the effects of helical motion on density fluctuations, we consider a simple situation in which all particles are driven by the same constant torque $\bm{\Omega}$.

Figure~\ref{fig1}(b) shows typical trajectories of a single free particle for several values of $D$, with time measured in units of $\tau = \sigma / v_0$, where $\sigma$ is the particle diameter.
The torque vector $\bm{\Omega}$ is oriented along the $z$-axis: $\bm{\Omega}=(0,0,\Omega)$.
When $D$ is very small, the trajectory becomes helical along the direction of $\bm{\Omega}$, provided that the $z$-component of the orientational vector $\bm{e}_j(t)$ is nonzero.
In the noiseless case ($D = 0$), the particle performs a perfect helical motion.
As $D$ increases, the trajectories become more stochastic and the influence of chirality becomes less pronounced.
The single-particle dynamics of 3D cABP have been characterized in Refs.~\cite{Sevilla2016PRE, Pattanayak_2024}, based on mean-square displacement and higher-order moments.

\begin{figure}[t]
\centering
\hspace*{-1.5mm}
  \includegraphics[width=8.8cm]{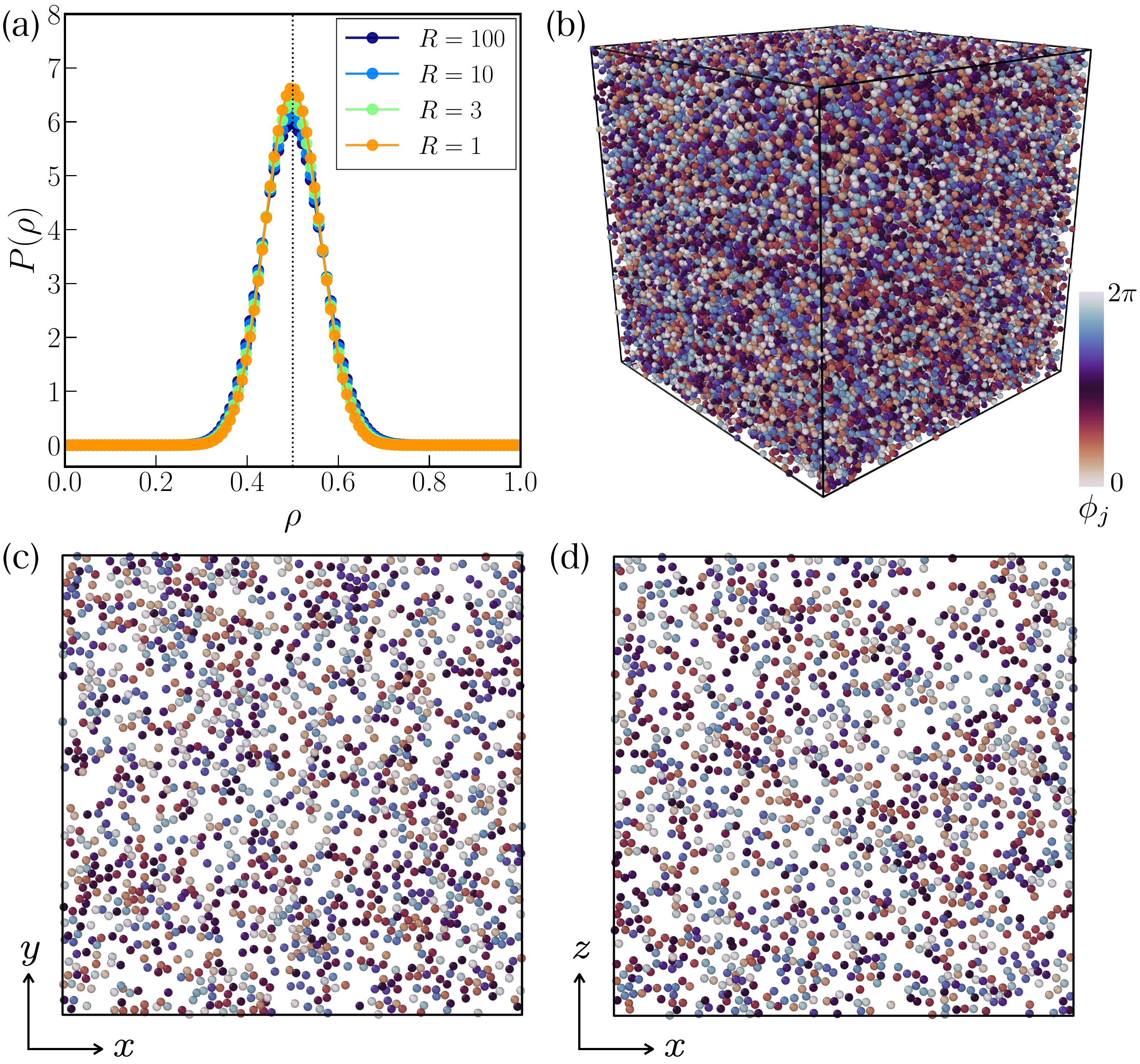}
  \caption{\label{fig2}(a)~The local density distribution $P(\rho)$ at $D = 0$.
In this figure, $\rho$ denotes the local density.
The vertical dotted line indicates the mean density $\rho = 0.5$.
(b)~A typical particle configuration in the steady state at $D = 0$, $\rho = 0.5$, and $R = 1$.
The color represents the azimuthal angle $\phi_j$ of the orientation vector $\bm{e}_j$, defined as $\bm{e}_j = (\sin\theta_j \cos\phi_j, \sin\theta_j \sin\phi_j, \cos\theta_j)$, where $\theta_j$ is the zenith angle.
Cross-sectional particle configurations with thickness $\sigma$ in (c)~the $x$-$y$ plane at $z = L/2$ and (d)~the $x$-$z$ plane at $y = L/2$.
}
\end{figure}
\begin{figure}[t]
\centering
\hspace*{-3.3mm}
  \includegraphics[width=8.8cm]{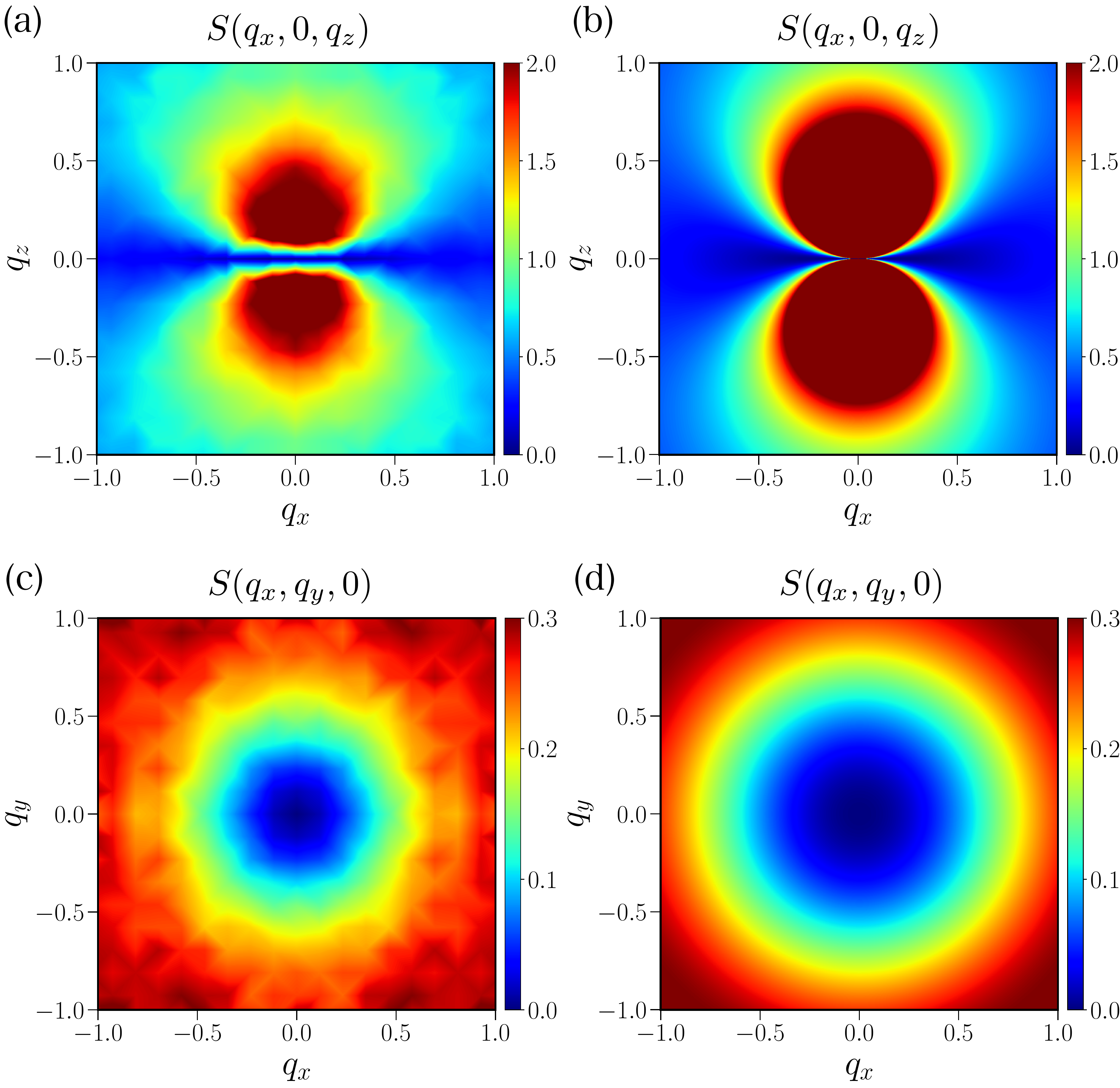}
  \caption{\label{fig3}Contour plots of the static structure factor $S(\bm{q})$ at $D = 0$ and $R = 1$ (or $\Omega = 1$).
The upper panels show $S(\bm{q})$ in the $q_x$-$q_z$ plane obtained from (a)~numerical simulations and (b)~theoretical analysis.
The lower panels show $S(\bm{q})$ in the $q_x$-$q_y$ plane obtained from (c)~numerical simulations and (d)~theoretical analysis.
}
\end{figure}
\begin{figure*}[t]
\centering
\hspace*{-3.3mm}
  \includegraphics[width=18cm]{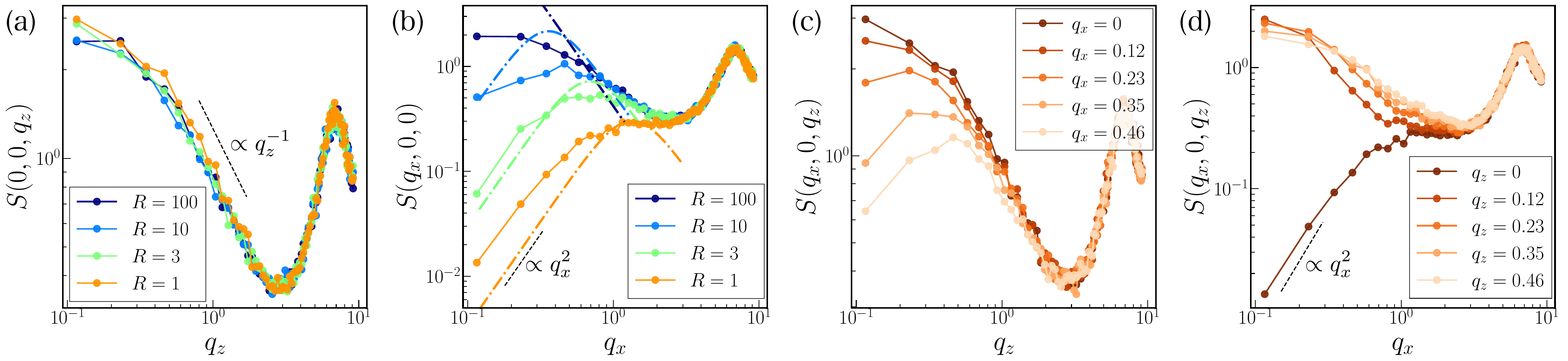}
  \caption{\label{fig4} The static structure factor $S(\bm q)$ obtained from the numerical simulation at $D=0$ on (a)~the $q_z$-axis and (b)~the $q_x$-axis. 
 The dashed-dotted lines in (b) represent the predictions of the linearized theory.
 (c) and (d) show the static structure factor $S(\bm q)$ in the $q_x$-$q_z$ plane at $R=1$ and $D = 0$. 
 (c)~$S(q_x,0,q_z)$ as a function of $q_z$ for various $q_x$. (d)~$S(q_x,0,q_z)$ as a function of $q_x$ for various $q_z$.
}
\end{figure*}
\section{Numerical simulation}
\label{sec3}
\subsection{Setup}
We numerically simulate a system of $N$ interacting particles, governed by Eqs.~(\ref{eq7-1}) and (\ref{eq7-2}), in a cubic box of side length $L$ with periodic boundary conditions.
The details of the simulation setup are as follows.
We numerically integrate the equation for the orientation, \eq{eq7-2}, by the Euler--Maruyama method.
The vector $\bm e_j(t)$ is normalized after each time step to ensure that the magnitude is unity. 
For the pairwise interaction, we consider only a short-range repulsive force, as in previous studies of the two-dimensional case~\cite{Ma2017, Li2019, Ma2022}. 
To this end, we employ the Weeks--Chandler--Andersen (WCA) potential given by~\cite{WCA}
\begin{equation}
U(r) = 4\epsilon\left[ \left( \frac{\sigma}{r }\right)^{12} - \left( \frac{\sigma}{r }\right)^{6} + \frac{1}{4} \right] \label{wca}
\end{equation}
for $r < 2^{1/6}\sigma$ and $U(r)=0$ otherwise.
Here, $\sigma$ is the diameter of a particle and $2^{1/6}\sigma$ is the cut-off length.
The units of length and time are chosen as $\sigma$ and $\tau = \sigma / v_0$, respectively. 
Unless otherwise noted, all parameters are expressed in these units; that is, we set $\sigma = 1$ and $v_0 = 1$.
The time step is set to $\Delta t = 10^{-3}$, and the number of particles is $N = 8 \times 10^4$.
Without loss of generality, we align the torque vector $\bm{\Omega}$ with the $z$-axis.
The control parameters are the mean number density $\rho$, the rotational diffusion constant $D$, the orbital radius $R = v_0 / \Omega$, and the dimensionless energy ratio $\mu \epsilon / (\sigma v_0)$.
Throughout this section, we focus on a low-density $\rho = 0.5$ (corresponding to a packing fraction of $N\pi\sigma^3/(6L^3) \simeq 0.26$), and we fix the parameter $\mu \epsilon / (\sigma v_0) = 1/24$. 
We compute the static structure factor $S(\bm q) = \expval*{\sum_{j,k=1}^{N}e^{-i\bm q\cdot (\bm r_j-\bm r_k)}}/N$ after confirming that the system has reached a stationary state by monitoring the time evolution of the potential energy.
We also verify that the system remains homogeneous and does not exhibit clustering or phase separation for any of the parameter sets used in this study, whereas the ABP model without chirality is known to undergo motility-induced phase separation (MIPS) even in three dimensions~\cite{Wysocki_2014, Stenhammar2014soft_matter, Farage2015PRE, Siebert2017}.
The effects of chirality on MIPS in three dimensions are beyond the scope of the present study.

\subsection{Results}
We first consider the noiseless limit ($D = 0$), in which a free particle exhibits a perfect helical motion.
Figure~\ref{fig2}(a) shows the local density distribution $P(\rho)$ for $D = 0$. 
Here, we calculate the local density by averaging the number of particles within small spherical regions of radius $3\sigma$.
For all values of $R$, the distribution is unimodal with a peak at the mean density, indicating that the system remains homogeneous.
Figure~\ref{fig2}(b) presents a representative snapshot of the simulation at $D = 0$, where the color positions the azimuthal angle $\phi_j$ of the orientation vector $\bm{e}_j$. 
We also present sliced snapshots of the particle configurations in the $x$-$y$ and $x$-$z$ planes, shown in Figs.~\ref{fig2}(c) and (d), respectively.
The particle positions and orientations are randomly distributed, and the system exhibits no characteristic structures such as clustering or synchronization.

We now observe the static structure factor $S(\bm q)$.
Note that $S(\bm q)$ cannot be reduced to a function of the scalar variable $q=|\bm q|$ as $S(q)$ because the system is not invariant under rotations about the $x$- or $y$-axes.

In Figs.~\ref{fig3}(a) and (c), we show the static structure factor at $R = 1$ in the $q_x$-$q_z$ plane and in the $q_x$-$q_y$ plane, respectively.
Reflecting the anisotropy of the system, the static structure factor in the $q_x$-$q_z$ plane exhibits an anisotropic pattern, whereas it remains isotropic in the $q_x$-$q_y$ plane.
To examine this anisotropy in more detail, we plot $S(0, 0, q_z)$ and $S(q_x, 0, 0)$ in Figs.~\ref{fig4}(a) and (b), respectively.
On the $q_z$-axis, $S(\bm{q})$ is enhanced at small $q_z$ and is insensitive to $R$ or chirality $\Omega$, which is natural since the $z$-component of \eq{eq7-2} does not depend on $\Omega$.
In contrast, $S(\bm{q})$ on the $q_x$-axis appears to vanish in the limit $q_x \rightarrow 0$, {\it i.e.}, hyperuniformity with the exponent $\alpha = 2$ is observed.
Although hyperuniformity is not evident for large radii, $R = 100$ and $10$, it is expected that $S(q_x, 0, 0)$ eventually goes to zero with the same hyperuniform exponent if the system size is sufficiently large.
These observations indicate that the static structure factor in this system is singular at the origin of Fourier space: $\lim_{q_z \rightarrow 0} S(0, 0, q_z) \neq \lim_{q_x \rightarrow 0} S(q_x, 0, 0)$.
Since the system is isotropic in the $x$-$y$ plane, more generally, $\lim_{q_z \rightarrow 0} S(0, 0, q_z) \neq \lim_{q \rightarrow 0} S(q_x, q_y, 0)$.
We also show the behavior of the static structure factor $S(q_x, 0, q_z)$ as a function of $q_z$ and $q_x$ with fixed $q_x, q_z (\geq 0)$ in Figs.~\ref{fig4}(c) and (d).
In Fig.\ref{fig4}(c), we observe that increasing $q_x$ leads to suppression of $S(q_x, 0, q_z)$ as a function of $q_z$ in the small-wavenumber regime.
In other words, $S(q_x, 0, q_z)$ is maximized at $q_x = 0$, which is also evident in \fig{fig3}(a).
$S(q_x, 0, q_z)$ as a function of $q_x$, shown in \fig{fig4}(d), exhibits hyperuniformity only at $q_z = 0$, while for $q_z > 0$, the density fluctuations are instead enhanced.

Anisotropic patterns and singularities in the static structure factor have also been observed in other anisotropic nonequilibrium systems. 
In Ref.~\cite{Nakano2024PRR}, the authors studied active Brownian particles with anisotropic self-propulsion in two dimensions and reported singular density correlations in the homogeneous fluid phase.
Lattice models of active fluids with spatial anisotropy are also known to exhibit similar features in density correlations, as shown by numerical simulations and field-theoretical analysis~\cite{Adachi2022PRR, Adachi2024PRR}.
In addition, passive anisotropic systems driven out of equilibrium, such as randomly driven diffusive gases, have long been known to exhibit singular density correlations~\cite{Schmittmann1995, Schmittmann1998}. 
The anisotropic pattern of $S(\bm q)$ in these systems is referred to as the ``butterfly'' or ``owl'' pattern~\cite{Schmittmann1998}. 
These observations suggest that the anisotropic pattern and singularity of $S(\bm q)$ are robust features of anisotropic nonequilibrium fluids. 
However, our system differs from the above examples in that the density fluctuations are hyperuniform in the plane perpendicular to the torque vector $\bm{\Omega}$. 
Such anisotropic hyperuniform states were previously proposed in Ref.~\cite{Torquato2016PRE}, and our system can be regarded as a concrete realization of this concept.

Qualitatively, the singular behavior of the density correlation in the system can be understood from established properties of standard ABP and cABP in two dimensions.
First, in the fluid phase of standard ABP, it is known that spatial velocity correlations develop~\cite{Caprini2020PRR, Szamel2021EPL}, and, concomitantly, long-wavelength density fluctuations are enhanced~\cite{kuroda2023, Szamel2024SoftMatter}.
These density fluctuations possess a finite correlation length given by $\xi \propto \sqrt{1/D}$, and become long-ranged in the limit $D = 0$ or infinite persistence time~\cite{kuroda2023, Szamel2024SoftMatter}.
Second, two-dimensional cABP systems are known to exhibit hyperuniformity, characterized by the exponent $\alpha = 2$, as confirmed by both numerical simulations~\cite{Li2019} and theoretical analysis~\cite{kuroda2023_chiral}.
Since the particle motion in our model can be effectively regarded as ``one-dimensional ABP" along the $z$-axis and two-dimensional cABP in the $x$-$y$ plane, it is natural that $S(0, 0, q_z)$ exhibits enhancement, whereas $S(q_x, q_y, 0)$ displays hyperuniformity.

\begin{figure}[t]
\centering
\hspace*{-3.3mm}
  \includegraphics[width=9cm]{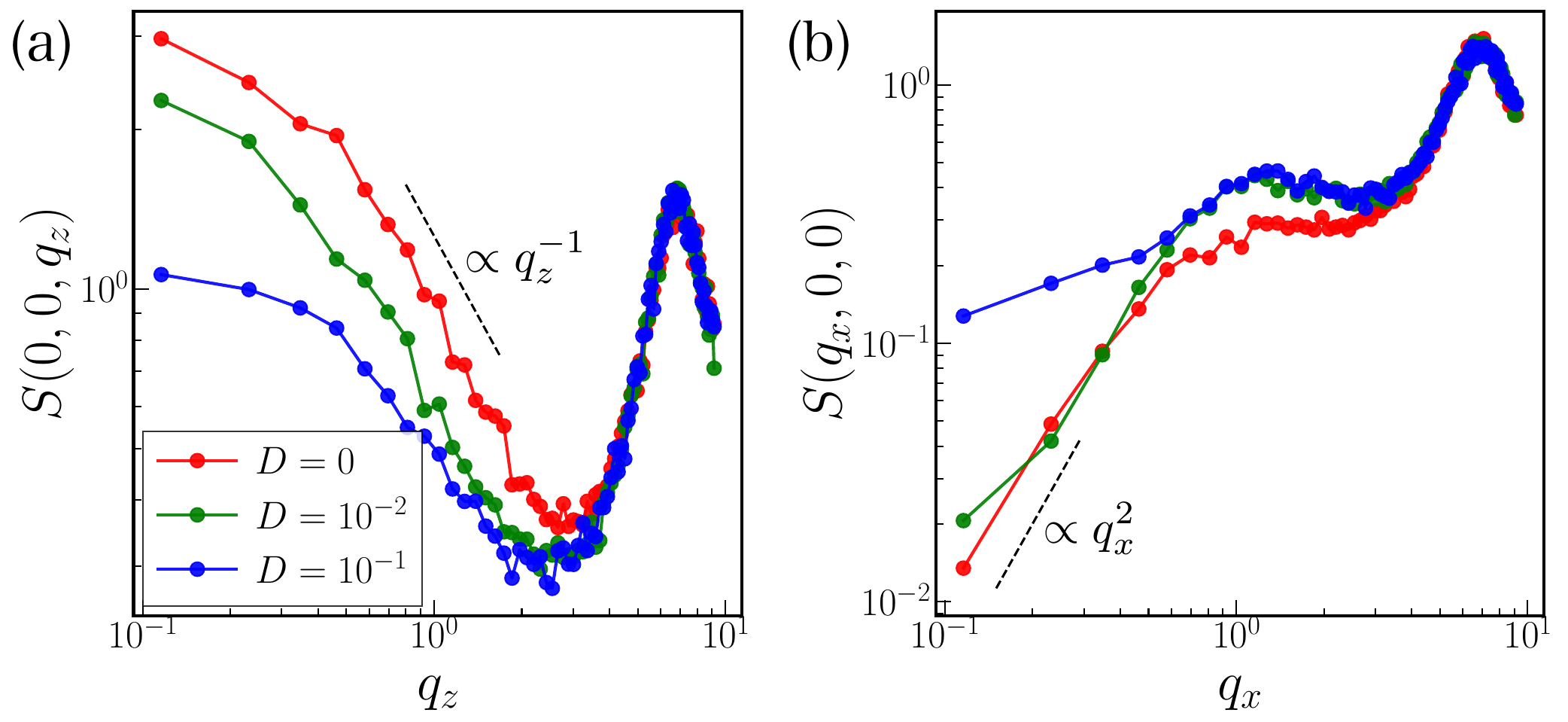}
  \caption{\label{fig5} Static structure factor obtained from the numerical simulation at $\rho=0.5$ and $R =1$. Panels (a) and (b) show the static structure factor on the $q_z$ and $q_x$-axis for various $D$, respectively. 
 In panel (a), the black dashed line represents $q_z^{-1}$ as a guide for the eye.
 The black dashed line in panel (b) is proportional to $q_x^{2}$. 
}
\end{figure}
We finally examine the effect of the rotational diffusion constant $D$ (or the persistence time) on the density correlation function.
Figure~\ref{fig5}(a) shows the static structure factor along the $q_z$-axis, $S(0, 0, q_z)$, for several values of $D$ at $R = 1$.
As $D$ increases, the intensity of $S(0, 0, q_z)$ at small $q_z$ decreases.
This trend is consistent with the fact that our model approaches equilibrium in the limit $D \to \infty$.
In Fig.~\ref{fig5}(b), we show the static structure factor along the $q_x$-axis, $S(q_x, 0, 0)$, for various values of $D$.
As $D$ increases, the hyperuniform behavior appears to vanish.
Although we cannot definitively conclude the disappearance of hyperuniformity for finite $D$ due to the limited system size, it is natural to infer that $S(q_x \to 0, 0, 0)$ remains nonzero for $D \neq 0$, similar to the case of two-dimensional cABP, which exhibit perfect hyperuniformity only when $D = 0$~\cite{Li2019, kuroda2023_chiral}.
Furthermore, by comparing the numerical values of $S(0, 0, q_z)$ and $S(q_x, 0, 0)$ at the smallest wavenumber, we find that the density correlation remains singular even at finite $D$.

\section{Theory}
\label{sec4}
To understand the numerical results presented in the previous section, we develop a fluctuating hydrodynamic theory for the homogeneous fluid state of 3D cABP and calculate the static structure factor through linear analysis.

\subsection{Fluctuating hydrodynamic equations}
Our interest lies in long-wavelength density fluctuations.
To analyze such large-scale behavior of the density field, hydrodynamic descriptions provide a standard theoretical tool.
Several approaches have been proposed to derive hydrodynamic equations for active particle models.
A prevalent method involves deriving equations for the average density and polarization by reducing the Fokker--Planck equation to an effective one-body equation and applying closure approximations~\cite{Bialk2013, Speck2015, Vrugt2023}.
This approach has also been applied to chiral active particles in two dimensions to investigate instabilities~\cite{Li2019, Bickmann2022, Ma2022, Sansa2022, Kreienkamp_2022}.
Another approach is based on kinetic theories for active matter, which has recently been considered for chiral active particles~\cite{Kalz:2024aa}.
However, in these approaches, noise terms must be added by hand when one wants to describe fluctuations of hydrodynamic variables~\cite{marconi2021, Speck2022PRE}.
To address fluctuations, we adopt Dean's method~\cite{Dean_1996, Nakamura_2009, illien2025}, which enables the derivation of hydrodynamic equations with noise terms.
In the case of two-dimensional cABP, this method successfully reproduces hyperuniformity with the same exponent as observed in numerical simulations, although discrepancies remain in the amplitude of the density correlation function~\cite{kuroda2023_chiral}.

The field variables that we consider are the density field and polarization, defined by 
\begin{align}
\rho(\bm r,t) &= \sum_{j=1}^{N}\delta(\bm r -\bm r_j(t)),  \label{density} \\
\bm p(\bm r,t) &= \sum_{j=1}^{N}\bm e_j(t) \delta(\bm r -\bm r_j(t)), \label{pola}
\end{align}
respectively. 
By employing Dean's method~\cite{Dean_1996, Nakamura_2009}, one can find the following set of equations for the density field $\rho(\bm r,t)$ and the polarization $\bm p(\bm r,t)$ (see Appendix~\ref{apeA} for the derivation):
\begin{align}
\p_{t} {\rho(\bm r,t)} &= -\nabla\cdot \bm J(\bm r,t), \label{eq7-3} \\
\bm J(\bm r,t) &= -\mu \nabla\cdot \mathsf P(\bm r,t) + v_{0}\bm p(\bm r,t), \label{eq7-4} \\
\p_{t}{\bm p(\bm r,t)} &= - \nabla\cdot\qty( \frac{\bm J(\bm r,t) \bm p(\bm r,t)}{\rho(\bm r,t)}) -2D\bm p(\bm r,t)  \notag \\
&\ \ \ + \bm \Omega \times \bm p(\bm r,t) +\sqrt{\frac{4D\rho(\bm r,t)}{3}}\bm \Upsilon(\bm r,t),
\label{eq7-6}
\end{align}
where the first term on the RHS of \eq{eq7-4} is the ``pressure'' tensor defined as 
\begin{equation}
\nabla\cdot \mathsf P(\bm r) = \rho(\bm r) \nabla \fdv{\mathcal F[\rho(\cdot)]}{\rho(\bm r)} \label{pressure}
\end{equation}
with
\begin{equation}
\mathcal F[\rho(\cdot)] = \frac{1}{2}\int_{V}\dd[3]\bm r\int_{V}\dd[3]\bm r'\ \rho(\bm r)\rho(\bm r')U(|\bm r-\bm r'|). \label{free_ene}
\end{equation}
The explicit form of the pressure tensor in terms of the density field and the pair interaction is given by the Irving--Kirkwood formula~\cite{Irving_Kirkwood, Kruger2018JChemPhys}, but we do not need it in the following discussion. 
$\bm \Upsilon(\bm r,t)$ in \eq{eq7-6} is a Gaussian white noise with zero mean $\expval{\Upsilon_{\alpha}(\bm r,t)}=0$ and the correlation $\expval{\Upsilon_{\alpha}(\bm r,t) \Upsilon_{\beta}(\bm r',t')} = \delta_{\alpha,\beta}\delta(\bm r-\bm r')\delta(t-t')$.
Equations~(\ref{eq7-3})-(\ref{eq7-6}) have the same form as that in the two-dimensional case~\cite{kuroda2023_chiral} except for the coefficients in \eq{eq7-6}.
We remark that Eqs.~(\ref{eq7-3})-(\ref{eq7-6}) reduce to the Dean--Kawasaki equation for overdamped equilibrium systems in the limit $D\rightarrow 0$ with $T_\RM{act}= v_0^2/(6D\mu)$ being constant~\cite{kuroda2023_chiral}.

Equations~(\ref{eq7-3})-(\ref{eq7-6}), derived via an extended Dean's method, are rigorous except for an assumption that each particle does not occupy the same position simultaneously (see Appendix~\ref{apeA}).
Therefore, Eqs.~(\ref{eq7-3})-(\ref{eq7-6}) have the same information as the original equations, Eqs.~(\ref{eq7-1}) and (\ref{eq7-2}). 
We henceforth focus on the large-scale or low-wavenumber limit. 
Since in homogeneous fluid states at the coarse-grained scales, one expects higher-order terms in gradients and in density fluctuations to be negligible, one assumes that the pressure gradient can be written as 
\begin{align}
\nabla\cdot \mathsf P(\bm r) \simeq \frac{1}{\rho\chi}\nabla \delta\rho(\bm r), \label{assumption}
\end{align}
where the coefficient $\chi$ is defined by $\chi^{-1} = \rho\eval{\p P/\p\rho}_{\rho(\bm r)=\rho}$ with $P(\bm r) = \RM{Tr}[\mathsf P(\bm r)]/3$. 
Although this assumption is crude, it is enough to capture the qualitative behavior of the density correlation function, as we will see below. 
If we wish to study instabilities or nonlinear effects, the pressure tensor needs to be further expanded to higher-order terms in $\rho$ and $\bm p$~\cite{Speck2022PRE}, but it is outside the scope of the present study. 
Together with the assumption, \eq{assumption}, and the linearization of equations around the homogeneous solutions, we have
\begin{align}
&\p_t \delta\rho(\bm r,t) = b \nabla^2 \delta\rho(\bm r,t)  - v_0 \nabla\cdot \delta\bm p(\bm r,t) , \label{eq7-7}  \\
&\p_{t} \delta\bm p(\bm r,t) = - 2D\delta\bm p(\bm r,t)  + \bm \Omega \times \delta\bm p(\bm r,t)  + \sqrt{\frac{4D\rho}{3}}\bm \Upsilon(\bm r,t),\label{eq7-8}
\end{align}
where we defined $b: = \mu/(\rho\chi)$.

Before looking at the density correlation function, we remark on the equilibrium limit of the linearized equations, Eqs.~(\ref{eq7-7}) and (\ref{eq7-8}).
Since \eq{eq7-8} is closed for the polarization, $\bm \xi(\bm r,t):= v_0\delta \bm p(\bm r,t)$ can be regarded as a colored noise.
This colored noise $\bm \xi(\bm r,t)$ has zero mean and the correlation 
\begin{equation}
\expval{\bm \xi(\bm r,t)\bm \xi^\RM{T}(\bm r',t')} = \frac{v_0^2 \rho}{3}\mathsf R_{\bm \Omega}(\Omega(t-t'))e^{-2D|t-t'|}\delta(\bm r-\bm r'), \label{act_noise}
\end{equation}
where $\RM{T}$ denotes the transpose.
$\mathsf R_{\bm \Omega}(\theta)$ in \eq{act_noise} is the rotation matrix representing a rotation by an angle $\theta$ around the vector $\bm \Omega$, given by the Rodrigues' formula:
\begin{align}
\mathsf R_{\bm \Omega}(\theta) = e^{\theta \mathsf K(\bm \Omega)},\ \ \ 
K_{\alpha,\beta} (\bm \Omega)= \frac{1}{\Omega}\sum_{\mu\in\{x,y,z\}}\epsilon_{\alpha\mu\beta}\Omega_\mu,
\end{align}
where $\Omega = | \bm \Omega|$ and $\epsilon_{\alpha\beta\gamma}$ is the Levi--Civita tensor.
In the limit $D\rightarrow \infty$ while keeping $T_\RM{act}= v_0^2/(6D\mu)$ constant, \eq{act_noise} becomes a Gaussian white noise and the fluctuation-dissipation relation is recovered:
\begin{equation}
\expval{\bm \xi(\bm r,t)\bm \xi^\RM{T}(\bm r',t')} = 2\mu T_\RM{act}\rho\delta(t-t')\delta(\bm r-\bm r')\mathbbm{1},
\end{equation}
where $\mathbbm{1}$ is the identity matrix. 
\subsection{Static structure factor}
We now calculate the equal time density correlation function in Fourier space, the static structure factor, defined by 
\begin{equation}
S(\bm q) = \frac{1}{N}\expval{\delta\tilde{\rho}(\bm q, 0)\delta \tilde{\rho}^*(\bm q, 0)}. \label{Sq_def}
\end{equation}
where $*$ represents the complex conjugate, and $\tilde X(\bm q,t)$ denotes the Fourier transform of $X(\bm r,t)$ with respect to $\bm r$: $\tilde X(\bm q,t)= \int_{V}\dd[3]\bm r\ X(\bm r,t)e^{-i\bm q\cdot \bm r}$. 
We also write the spatiotemporal Fourier transform of $X(\bm r,t)$ as $\hat X(\bm q,\omega) = \int_{-\infty}^{\infty}\dd t\ \tilde X(\bm q,t)e^{i\omega t}$.
In Fourier space, Eqs.~(\ref{eq7-7}) and (\ref{eq7-8}) read
\begin{equation}
\mathsf G^{-1}(\bm q,\omega) \hat{\bm \Phi}(\bm q,\omega) = \hat{\bm \Xi}(\bm q,\omega),
\end{equation}
where we defined $\bm \Phi(\bm r,t) = (\rho(\bm r,t), \bm p(\bm r,t))$, $\bm \Xi(\bm r,t) = \sqrt{4D\rho/3}(0,\bm \Upsilon(\bm r,t))$, and 
\begin{align}
&\mathsf G^{-1}(\bm q,\omega)  \notag \\  &= 
\mqty(
- i\omega + bq^2 & iv_0 q_x & iv_0 q_y & iv_0 q_z\\
0 & -i\omega + 2D& \Omega_z & -\Omega_y \\
0 & -\Omega_z & -i\omega + 2D & \Omega_x\\
0 & \Omega_y & -\Omega_x & -i\omega + 2D
). \label{15}
\end{align}
The dynamical correlation functions in $(\bm q,\omega)$-space can be calculated as
\begin{align}
{S}_{a,b}(\bm q,\omega) 
&:= \frac{1}{N}\int_{-\infty}^{\infty}\dd t\ \expval{ \tilde{\Phi}_a(\bm q,t)\tilde{\Phi}_b^*(\bm q,0)}e^{i\omega t} \notag \\
& = \frac{4D}{3}\sum _{c=2}^{4}G_{a,c}(\bm q,\omega) G_{b,c}^*(\bm q,\omega).
\end{align}
Using this formula, the dynamical structure factor, $S(\bm q,\omega):=S_{1,1}(\bm q,\omega)$, ends up with
\begin{widetext}
\begin{equation}
S(\bm q,\omega ) = \frac{4v_0^2D}{3}\frac{q^2\omega^4 + \qty[(8D^2 +\Omega^2)q^2 - 3(\bm \Omega\cdot \bm q)^2]\omega^2 + (4D^2+\Omega^2)
\qty[4D^2q^2 + (\bm \Omega\cdot \bm q)^2 ]}
{
\qty(\omega^2 + b^2q^4)(\omega^2+4D^2)\qty[ \qty{ \omega^2 - (4D^2 + \Omega^2)}^2 + 16D^2\omega^2]
}. \label{eq7-16}
\end{equation}
The static structure factor, \eq{Sq_def}, is obtained by integrating \eq{eq7-16} over $\omega$ as
\begin{align}
S(\bm q) &= \frac{1}{2\pi}\int_{-\infty}^{\infty}\dd\omega \ S(\bm q,\omega) = \frac{v_0^2}{3} \frac{q^2(2D + bq^2)^2 + (\bm \Omega\cdot \bm q)^2}{bq^2(b q^2 + 2D)\qty[(bq^2+2D)^2 + \Omega^2]}. \label{eq7-17}
\end{align}
\end{widetext}

We first discuss the limit $D\rightarrow 0$, or equivalently, the infinite persistence time limit $\tau_\RM{p}=1/(2D)\rightarrow \infty$. 
Equation~(\ref{eq7-17}) in this limit reduces to
\begin{align}
S(\bm q) = \frac{v_0^2}{3}\frac{b^2q^6 + \Omega^{2}q_{z}^{2}}{b^2 q^4(b^2 q^4 + \Omega^2)} ,\label{eq7-18}
\end{align}
where the torque vector was chosen to be parallel to the $z$-direction as $\bm \Omega = (0,0,\Omega)$. 
Figures~\ref{fig3}(b) and (d) depict contour plots of \eq{eq7-18} on the $q_x$-$q_z$ and $q_x$-$q_y$ planes, respectively.
The static structure factor exhibits an anisotropic shape in the $q_x$-$q_z$ plane, while it appears isotropic in the $q_x$-$q_y$ plane.
These behaviors are qualitatively in agreement with the numerical observations shown in Figs.~\ref{fig3}(a) and (c). 
Furthermore, $S(\bm q)$ has a singularity at the origin, as indicated by the numerical simulation in the previous section. 
To see the singularity, we focus on the behavior on the $q_z$ and $ q_x$-axes.
Along the $q_z$-axis, \eq{eq7-18} reduces to
\begin{equation}
S(0,0,q_z) = \frac{v_0^2}{3b^2}\frac{1}{q_z^2}, \label{eq7-19}
\end{equation}
meaning that the density fluctuations increase at large scales for the $z$-direction. 
In contrast,, along the $q_x$-axis, $S(\bm q)$ becomes
\begin{equation}
S(q_x,0,0) = \frac{v_0^2}{3} \frac{q_x^2 }{ \Omega^2 + b^2q_x^4 }
= \frac{1}{3}(R q_x)^2 + O(q_x^6), \label{eq7-20}
\end{equation} 
where $R=v_0/\Omega$.
Equation~(\ref{eq7-20}) is a scaling of hyperuniformity with the exponent $\alpha=2$, which is identical to that of two-dimensional cABP~\cite{Li2019, kuroda2023_chiral}.
Therefore, \eq{eq7-18} is singular at the origin:
$\lim_{q_z\rightarrow 0} S(0,0,q_z) \neq \lim_{q_x\rightarrow 0} S(q_x,0,0)$. 
This singular behavior is also consistent with the numerical observations. 
However, the exponent of $S(0,0,q_z)$ disagrees with the numerical result. 
The static structure factor $S(0,0,q_z)$ obtained from the numerical simulation shown in \fig{fig4}(a) is $S(0,0,q_z)\sim 1/q_z$, in contrast to the theoretical prediction $S(0,0,q_z)\propto 1/q_z^{2}$.
Moreover, this scaling appears to terminate around $q_z = 0.4$, suggesting the existence of a characteristic length scale, whereas the theory predicts a scale-free behavior. 
We remark that the characteristic length scale observed in the numerically obtained $S(0,0,q_z)$ in Fig.~\ref{fig4}(a) is not attributable to finite-size structures, as there is no sign of clustering or phase separation (see Fig.~\ref{fig2}). 
Since our particle model can be effectively regarded as “one-dimensional ABP” along the $z$-axis, one possible reason for this discrepancy is the presence of anomalies specific to one-dimensional systems, such as single-file diffusion~\cite{Akintunde2025JChemPhys}, which are not captured by our theory.
Another possibility is the different behavior of the correlation length in one-dimensional ABP.
It has been reported that, in a narrow annular confinement, the correlation length saturates to a constant value as the persistence time $\tau_\RM{p}$ increases~\cite{Caprini2021JChemPhys_confined}, whereas the correlation length in two dimensions is proportional to $\sqrt{\tau_\RM{p}}$~\cite{Caprini2020PRR, Szamel2021EPL, kuroda2023}. 
This effect is also beyond the scope of our theory.
 In contrast, the theoretical prediction of the hyperuniform exponent in $S(q_x,0,0)$ agrees with the numerical data.  
However, the theoretical expression for $S(q_x,0,0)$, \eq{eq7-20}, is not fully quantitative, that is, a discrepancy in the prefactor remains.
The dash-dotted lines in \fig{fig4}(b) depict the theoretical predictions \eq{eq7-20}. 
The parameter $b$ is determined by the fitting but does not affect the hyperuniform scaling, as is obvious in the second equation in \eq{eq7-20}.
The theoretical line for $R=1$ deviates from the numerical data, while for $R=3$ the agreement is somewhat better.
As for $R=10$ and $100$, we cannot discuss quantitative assessment with the current system size. 
Further theoretical calculations, such as non-linear analysis, would be required for a quantitative description.

Finally, we discuss the case of finite rotational diffusion constants $D$.
On the $q_z$ and $q_x$-axis, \eq{eq7-18} is written as
\begin{align}
S(0,0,q_z) &= \frac{S_0}{1+(\xi q_z)^2},  \label{eq7-21}\\
S(q_x,0,0) &= \frac{v_0^2}{3b}\frac{2D+bq_x^2}{\Omega^2 + (bq_x^2 + 2D)^2}, \label{eq7-22}
\end{align}
respectively.
Here $S_0:= v_0^2/(6bD)$ and $\xi:=\sqrt{b/(2D)}$. 
Both on the $q_z$ and $q_x$-axis, $S(\bm q \rightarrow \bm 0)$ remains constant, but $S(\bm q)$ is still singular at the origin.
To measure how strong the singularity is, depending on the parameters $D$ and $\Omega$, we introduce the following ratio \cite{Schmittmann1998}: 
\begin{equation}
\mathcal R:= \frac{\lim_{q_z\rightarrow 0} S(0,0,q_z)}{ \lim_{q_x\rightarrow 0} S(q_x,0,0)} = 1+ \frac{\Omega^2}{4D^2} = 1 + (\Omega \tau_\RM{p})^2, \label{eq7-23}
\end{equation}
where we used $\tau_\RM{p}=1/(2D)$ in the last equality. 
This ratio becomes unity either when $\Omega = 0$ or in the limit $D \to \infty$.
In the former case, the system reduces to three-dimensional ABP without chirality~\cite{Wysocki_2014, Stenhammar2014soft_matter, Farage2015PRE, Siebert2017}, and the static structure factor takes the Ornstein--Zernike form $S(q) = {S_0}/[{1+(\xi q)^2}]$.
In the latter case, corresponding to the equilibrium limit, the structure factor becomes a constant, $S(q) = S_0 = \rho T_{\mathrm{act}} \chi$, which is a well-known expression of $S(q \to 0)$ in equilibrium statistical mechanics~\cite{hansen}.
Except in these two cases, $\mathcal R$ is not unity. 
Increasing $\Omega$ or decreasing $D$ renders the singularity stronger. 
In particular, $\mathcal R$ diverges in the limit $D\rightarrow 0$ or $\tau_\RM{p}\rightarrow \infty$.
This behavior stands in contrast to anisotropic active fluids and driven lattice gases, in which $\mathcal R$ remains finite~\cite{Schmittmann1995, Schmittmann1998, Adachi2022PRR, Nakano2024PRR}.
\section{Summary}
\label{sec5}
In this paper, we have studied density fluctuations in a three-dimensional chiral active fluid using a helical self-propelled particle model. 
We have considered a simple situation where all the particles have the same torque and perform a helical motion with a persistence time. 
Extensive numerical simulations of the model have revealed that the static structure factor in homogeneous states shows an anisotropic pattern and is singular at the origin in Fourier space. 
These features resemble those observed in other anisotropic nonequilibrium fluids, such as anisotropic active fluids~\cite{Adachi2022PRR, Nakano2024PRR, Adachi2024PRR} and driven diffusive gases~\cite{Schmittmann1995, Schmittmann1998}.
Moreover, in the limit of large persistence time, the system exhibits hyperuniformity in the direction perpendicular to the torque, while showing giant density fluctuations in the direction parallel to it.
To understand the numerical observations, we have developed an effective fluctuating hydrodynamic theory for homogeneous fluid states of 3D cABP.  
The theory successfully reproduces the same qualitative behavior of the density correlation functions as observed in numerical simulations, although it does not achieve full quantitative agreement.

As a future direction, it is important to investigate the full phase behavior of three-dimensional chiral active particles.
In two dimensions, chiral active particles are known to exhibit clustering~\cite{Li2019, Liebchen2017, Ma2022, Bickmann2022}, and it is therefore natural to expect similar behavior in three dimensions.
Moreover, two-dimensional chiral active particles can crystallize with long-range translational order~\cite{kuroda2025PRR}, suggesting that chirality may also play a significant role in crystallization in three dimensions.
In addition, while the present study assumes identical torques for all particles, exploring the effects of torque dispersion would be an important extension.

From an experimental perspective, singular density correlations found in the present study may be challenging to observe but could become accessible in future experiments, especially given that hyperuniformity in two-dimensional chiral active fluids has already been observed experimentally using swimming marine algae~\cite{Huang2021} and pear-shaped Quincke rollers~\cite{Zhang2022}.
For example, the marine algae used in Ref.~\cite{Huang2021} might also be suitable for studying the three-dimensional case, as they are known to exhibit helical trajectories when they are far from interfaces.
Additionally, bottom-heavy Janus particles swimming in three-dimensional space~\cite{Campbell:2013aa, Campbell2017JChemPhys} and magnetically driven particles~\cite{Chen:2025aa} could serve as potential experimental systems for investigating density correlations in three-dimensional chiral active fluids.
\begin{acknowledgments}
Y.K. acknowledges support by JSPS KAKENHI (grant no.~JP23KJ1068).
T.K. acknowledges support by the JST FOREST Program (grant no.~JPMJFR212T), AMED Moonshot Program (grant no.~JP22zf0127009), JSPS KAKENHI (grant no.~JP24H02203), and Takeda Science Foundation.
K.M. acknowledges support by JSPS KAKENHI (grant no.~JP24H00192).
\end{acknowledgments}
\appendix
\section{Derivation of fluctuating hydrodynamic equations}
\label{apeA}
In this appendix, we derive the fluctuating hydrodynamic equations by extending Dean's method \cite{Dean_1996, Nakamura_2009} to 3D cABP.
\subsection{Derivation of Eqs.~(\ref{eq7-4})-(\ref{eq7-6})}
First, we derive the equation for the density field. The time derivative of \eq{density} leads to the continuity equation as
\begin{equation}
\p_t{\rho(\bm r,t)} = -\nabla\cdot \bm J(\bm r,t), \label{a1}
\end{equation}
where the density current $\bm J(\bm r,t)$ is given by
\begin{align}
\bm J(\bm r,t) &:= \sum_{j=1}^{N}\dot{ \bm r} _j(t) \delta(\bm r-\bm r_j(t)) \notag \\
&= \mu \sum_{j=1}^{N}\bm F_j(t) \delta(\bm r-\bm r_j(t))  + v_0\bm p(\bm r,t). \label{a2}
\end{align}
The first term of \eq{a2} can be rewritten as
\begin{align}
\sum_{j=1}^{N}\bm F_j(t)\delta(\bm r- \bm r_j(t)) 
&= - \rho(\bm r,t)\nabla\fdv{\mathcal F[\rho(\cdot,t)]}{\rho(\bm r,t)}. \label{a3}
\end{align}
The functional $\mathcal F[\rho]$ is defined by \eq{free_ene}, and \eq{a3} can be interpreted as the divergence of the  ``pressure'' tensor as in \eq{pressure}.
We next construct the equation for the polarization defined in \eq{pola}. 
Taking the time derivative of \eq{pola}, we get
\begin{equation}
\p_t{\bm p(\bm r,t)} = - \nabla\cdot \mathsf M(\bm r,t)  + \sum_{j=1}^{N}\dv{\bm e_j(t)}{t} \delta(\bm r -\bm r_j(t)), \label{a4}
\end{equation}
where the tensor $\mathsf M(\bm r,t)$ is defined as
\begin{equation}
\mathsf M(\bm r,t):=  \sum_{j=1}^{N}\dot{\bm r}_j(t)\bm e_j(t) \delta(\bm r -\bm r_j(t)). \label{a5}
\end{equation}
To express \eq{a5} in terms of the hydrodynamic fields, we assume that
\begin{equation}
\delta(\bm r_j(t) - \bm r_k(t)) = \delta(\bm r_j(t) - \bm r_k(t))\delta_{j,k}, \label{a6}
\end{equation}
following Ref.~\cite{Nakamura_2009}.
This relation is correct if the particles $j$ and $k$ never occupy the same position simultaneously, such as when the pairwise potential $U( r)$ is the repulsive interaction~\cite{Nakamura_2009}.
With the assumption \eq{a6}, \eq{a5} can be written as 
\begin{equation}
\mathsf M(\bm r,t) = \frac{\bm J(\bm r,t)\bm p(\bm r,t)}{\rho(\bm r,t)}. \label{a7}
\end{equation}
To rewrite the last term on the RHS of \eq{a4}, we use \eq{eq7-2} in the It\^o representaion~\cite{gardiner}:
\begin{equation}
\dv{ \bm e_j(t)}{t} = - 2D \bm e_j(t) + \qty( \bm \Omega + \sqrt{2D} \bm \eta_j(t) )\overset{\RM{i}}{\times}  \bm e_j(t) , \label{a8}
\end{equation}
where the symbol $\overset{\RM{i}}{\times}$ means that the cross product is taken in the It\^o sense.
By substituting Eqs.~(\ref{a7}) and (\ref{a8}) into \eq{a4}, we have 
\begin{align}
\p_t{\bm p(\bm r,t)} = - \nabla\cdot\qty( \frac{\bm J(\bm r,t) \bm p(\bm r,t)}{\rho(\bm r,t)}) -2D\bm p(\bm r,t)  \notag \\
+ \bm \Omega \times \bm p(\bm r,t) +\bm \Lambda(\bm r,t),
\end{align}
where $\bm \Lambda(\bm r,t)$ is defined by
\begin{equation}
\bm \Lambda(\bm r,t):=\sqrt{2D} \sum_{j=1}^{N} \bm \eta_j(t)\overset{\RM{i}}{\times}  \bm e_j(t)\delta(\bm r-\bm r_j(t)). \label{a10}
\end{equation}
Equation~(\ref{a10}) is not closed for the hydrodynamic fields but can be written as the following simple form in the large $N$ limit:
\begin{equation}
\bm \Lambda(\bm r,t) = \sqrt{ \frac{4D\rho(\bm r,t)}{3} } \bm \Upsilon(\bm r,t), \label{a11}
\end{equation}
where $\bm \Upsilon(\bm r,t)$ is a Gaussian white noise with zero mean and correlation $\expval{\Upsilon_{\alpha}(\bm r,t) \Upsilon_{\beta}(\bm r',t')} = \delta_{\alpha,\beta}\delta(\bm r-\bm r')\delta(t-t')$.
We thus reach Eqs.~(\ref{eq7-3})-(\ref{eq7-6}) aside from the derivation of \eq{a11}.

\subsection{Justification of \eq{a11}}
Here, we show that the expectation value and the two-point correlation of  \eq{a10} are identical to those of \eq{a11} in the large $N$ limit.
First, the expectation value of \eq{a10} is obviously zero since the cross product is taken in the It\^o representation.
Next, the correlation between the elements $\alpha$ and $\beta$ of \eq{a10} is
\begin{align}
\expval{\Lambda_\alpha(\bm r,t) \Lambda_\beta(\bm r',t') }
=2D\sum_{j=1}^{N}\delta(t-t')\delta(\bm r - \bm r') \delta(\bm r-\bm r_j(t)) \notag \\
\times \sum_{\mu,\nu,\sigma\in\{x,y,z\}}\epsilon_{\alpha\mu\nu}\epsilon_{\beta\mu\sigma}\expval{e_j^{(\nu)}(t)e_j^{(\sigma)}(t)}. \label{a12}
\end{align}
To estimate this correlation, we calculate the autocorrelation function $\expval*{e_j^{(\alpha)}(t)e_j^{(\beta)}(t)}$. 
Using the It\^o formula~\cite{gardiner}, one can derive the differential equation for this autocorrelation function:
\begin{align}
\dv{t}\expval{e^{(\alpha)}_j(t)e^{(\beta)}_j(t)}=-6D\qty[ \expval{ e^{(\alpha)}_j(t) e^{(\beta)}_j(t)}  - \frac{1}{3}\delta_{\alpha,\beta}] \notag\\
+\sum_{\mu,\nu}
\qty[ \epsilon_{\alpha \mu \nu} \expval {e^{(\beta)}_j(t) e^{(\nu)}_j(t)} + \epsilon_{\beta \mu \nu} \expval{ e^{(\alpha)}_j(t) e^{(\nu)}_j(t) } ] \Omega_\mu .  \label{a13}
\end{align}
By introducing a symmetric matrix
\begin{equation}
\mathsf Q_j(t) :=   \expval{\bm e_j(t)\bm e_j^\RM{T}(t)} - \frac{1}{3}\mathbbm{1} \label{Qtensor}
\end{equation}
and an antisymmetric matrix  
\begin{equation}
H_{\alpha,\beta}:= \sum_{\mu}\Omega_\mu \epsilon_{\alpha\mu\beta},
\end{equation}
\eq{a13} can be written as 
\begin{equation}
\dv{t}\mathsf Q_j(t) = - \zeta \mathsf Q_j(t)  + [\mathsf H ,\mathsf Q_j(t)], \label{a16}
\end{equation}
where $\zeta:= 6D$ and $[\mathsf X,\mathsf Y]:=\mathsf X\mathsf Y  -\mathsf Y\mathsf X$ denotes the commutator.
Equation~(\ref{a16}) is the same form as the Heisenberg equation with a dissipation in imaginary time.
The formal solution of  \eq{a16} is given by
\begin{equation}
\mathsf Q_j(t) = e^{-\zeta t} e^{t \mathsf H}\mathsf Q_j(0) e^{-t \mathsf H}. \label{solusion}
\end{equation}
We now define 
\begin{equation}
\mathcal Q^{(N)}_{\alpha,\beta}(\bm r,t) := \sum_{j=1}^{N} Q_j^{(\alpha,\beta)}(0)\delta(\bm r-\bm r_j(t)) \label{a21}
\end{equation}
and evaluate this quantity in the large $N$ limit.
Here, $Q_j^{(\alpha,\beta)}(0)$ is an element of the matrix $\mathsf Q_j(0)$ in \eq{solusion}.
We assume that the initial orientation $\bm e_j(0)$ is chosen uniformly at random on the unit sphere.
Then 
\begin{align}
\lim_{N\rightarrow \infty} \frac{1}{N} \sum_{j=1}^{N}Q_j^{(\alpha,\beta)}(0) 
&= \frac{1}{4\pi} \int_{\abs{\bm e}=1}\dd[3] \bm e\ \qty(e_\alpha e_\beta -\frac{1}{3}\delta_{\alpha,\beta}) \notag \\
&=0 . \label{a22}
\end{align}
To evaluate Eq.~(\ref{a21}), we expand it in a Fourier series:
\begin{equation}
\mathcal Q^{(N)}_{\alpha,\beta}(\bm r,t)  = \sum_{\bm n\in \mathbb{Z}^3} \widetilde {\mathcal Q}^{(N)}_{\alpha,\beta}(\bm n,t) e^{2\pi i\bm n\cdot \bm r/L} 
\end{equation}
where the Fourier coefficient is given by
\begin{equation}
\widetilde {\mathcal Q}^{(N)}_{\alpha,\beta}(\bm n,t)  = \frac{1}{L^3} \sum_{j=1}^{N} Q_j^{(\alpha,\beta)}(0) e^{-2\pi i\bm n\cdot \bm r_j/L} .
\end{equation}
Since $\sum_{j=1}^{N} Q_j^{(\alpha,\beta)}(0)=o(N)=o(L^3)$ from \eq{a22}, we have
\begin{equation}
\abs{\widetilde {\mathcal Q}^{(N)}_{\alpha,\beta}(\bm n,t)} \leq \frac{1}{L^3}\abs{\sum_{j=1}^{N} Q_j^{(\alpha,\beta)}(0)} = o(1).
\end{equation}
Thus, \eq{a21} converges to zero in the large $N$ limit:
\begin{equation}
\lim_{N\rightarrow \infty}\mathcal Q^{(N)}_{\alpha,\beta}(\bm r,t) = 0. \label{a26}
\end{equation}
Using Eqs.~(\ref{Qtensor}), (\ref{solusion}), and (\ref{a26}), we obtain
\begin{equation}
\sum_{j=1}^{N} \expval{e_j^{(\alpha)}(t)e_j^{(\beta)}(t)}\delta(\bm r-\bm r_j) = \frac{\rho(\bm r,t)}{3}\delta_{\alpha,\beta} +o(1)
\end{equation}
in the limit $N\rightarrow \infty$. 
Therefore, \eq{a12} in the limit $N\rightarrow \infty$ reads
\begin{equation}
\expval{\Lambda_\alpha(\bm r,t) \Lambda_\beta(\bm r',t') } = \frac{4D}{3}\rho(\bm r,t)\delta_{\alpha,\beta}\delta(\bm r-\bm r')\delta(t-t') + o(1).
\end{equation}
This means that $\bm \Lambda(\bm r,t)$ in Eqs.~(\ref{a10}) and (\ref{a11}) is statistically identical, at least up to the second cumulant. 

%

\end{document}